\documentclass[aip,rsi,reprint,graphicx]{revtex4-1} 
\usepackage{amsmath,amssymb}
\usepackage{bm}
\usepackage{graphicx}
\usepackage{amsmath}
\usepackage{multirow}
\usepackage{threeparttable}
\usepackage{tabularx}
\usepackage{booktabs}

\draft 

\begin{document}


\title{Sine wave gating Silicon single-photon detectors for multiphoton entanglement experiments} 


\author{Nan Zhou}
\affiliation{Hefei National Laboratory for Physical Sciences at the Microscale and Department
of Modern Physics, University of Science and Technology of China, Hefei, Anhui 230026, China}
\affiliation{CAS Center for Excellence and Synergetic Innovation Center in Quantum Information
and Quantum Physics, University of Science and Technology of China, Hefei, Anhui 230026, China}

\author{Wen-Hao Jiang}
\affiliation{Hefei National Laboratory for Physical Sciences at the Microscale and Department
of Modern Physics, University of Science and Technology of China, Hefei, Anhui 230026, China}
\affiliation{CAS Center for Excellence and Synergetic Innovation Center in Quantum Information
and Quantum Physics, University of Science and Technology of China, Hefei, Anhui 230026, China}

\author{Luo-Kan Chen}
\affiliation{Hefei National Laboratory for Physical Sciences at the Microscale and Department
of Modern Physics, University of Science and Technology of China, Hefei, Anhui 230026, China}
\affiliation{CAS Center for Excellence and Synergetic Innovation Center in Quantum Information
and Quantum Physics, University of Science and Technology of China, Hefei, Anhui 230026, China}

\author{Yu-Qiang Fang}
\affiliation{Hefei National Laboratory for Physical Sciences at the Microscale and Department
of Modern Physics, University of Science and Technology of China, Hefei, Anhui 230026, China}
\affiliation{CAS Center for Excellence and Synergetic Innovation Center in Quantum Information
and Quantum Physics, University of Science and Technology of China, Hefei, Anhui 230026, China}

\author{Zheng-Da Li}
\affiliation{Hefei National Laboratory for Physical Sciences at the Microscale and Department
of Modern Physics, University of Science and Technology of China, Hefei, Anhui 230026, China}
\affiliation{CAS Center for Excellence and Synergetic Innovation Center in Quantum Information
and Quantum Physics, University of Science and Technology of China, Hefei, Anhui 230026, China}

\author{Hao Liang}
\affiliation{Hefei National Laboratory for Physical Sciences at the Microscale and Department
of Modern Physics, University of Science and Technology of China, Hefei, Anhui 230026, China}
\affiliation{CAS Center for Excellence and Synergetic Innovation Center in Quantum Information
and Quantum Physics, University of Science and Technology of China, Hefei, Anhui 230026, China}

\author{Yu-Ao Chen}
\affiliation{Hefei National Laboratory for Physical Sciences at the Microscale and Department
of Modern Physics, University of Science and Technology of China, Hefei, Anhui 230026, China}
\affiliation{CAS Center for Excellence and Synergetic Innovation Center in Quantum Information
and Quantum Physics, University of Science and Technology of China, Hefei, Anhui 230026, China}

\author{Jun Zhang}
\email{zhangjun@ustc.edu.cn}
\affiliation{Hefei National Laboratory for Physical Sciences at the Microscale and Department
of Modern Physics, University of Science and Technology of China, Hefei, Anhui 230026, China}
\affiliation{CAS Center for Excellence and Synergetic Innovation Center in Quantum Information
and Quantum Physics, University of Science and Technology of China, Hefei, Anhui 230026, China}

\author{Jian-Wei Pan}
\affiliation{Hefei National Laboratory for Physical Sciences at the Microscale and Department
of Modern Physics, University of Science and Technology of China, Hefei, Anhui 230026, China}
\affiliation{CAS Center for Excellence and Synergetic Innovation Center in Quantum Information
and Quantum Physics, University of Science and Technology of China, Hefei, Anhui 230026, China}

\date{\today}


\begin{abstract}
Silicon single-photon detectors (SPDs) are the key devices for detecting single photons in the visible wavelength range.
Here we present high detection efficiency silicon SPDs dedicated to the generation of multiphoton entanglement based on the technique
of high-frequency sine wave gating. The silicon single-photon avalanche diodes (SPADs) components are acquired by disassembling
6 commercial single-photon counting modules (SPCMs).
Using the new quenching electronics, the average detection efficiency of SPDs is increased from 68.6\% to 73.1\% at a wavelength of 785 nm.
These sine wave gating SPDs are then applied in a four-photon entanglement experiment, and the four-fold coincidence count rate is increased by 30\% without degrading its visibility compared with the original SPCMs.
\end{abstract}


\maketitle 

\section{introduction}

Silicon single-photon detectors (SPDs) provide the primary solution for the detection of ultraweak light in the visible wavelength range~\cite{EFM11}, and
are extensively used in diverse fields such as biology, lidar, quantum communication, and quantum information processing. A silicon SPD is composed of
a single-photon avalanche diode (SPAD) component and corresponding quenching circuit~\cite{Cova96,KJS11}.
Due to the intrinsic high performance of silicon SPAD and optimized quenching electronics, silicon SPD can exhibit much higher photon detection efficiency (PDE)
and much lower dark count rate (DCR) than photomultiplier tube.

So far, there are two kinds of commercial silicon SPDs, including single-photon counting modules (Excelitas)~\cite{SPCM} and PDM photon counting modules (Micro Photon Devices)~\cite{MPD}.
Single-photon counting modules (SPCMs) exploit silicon SPADs of patented SLiK~\cite{SLiK93,Cova04} structure with pretty thick depletion layer and 180 $\mu$m diameter active area,
and exhibit excellent performance with $\sim$ 70\% PDE at 650 nm, $\sim$ 100 cps DCR (depends on the part number), and $\sim$ 400 ps timing resolution. PDM photon counting modules, however, use epitaxial silicon SPADs with thin depletion layer ($\sim$ 1 $\mu$m) and small active area, and patented integrated active quenching circuits. These
SPDs exhibit $\sim$ 35\% PDE at 650 nm, $\sim$ 50 cps DCR (depends on the grade), and $\sim$ 50 ps timing resolution.

Nevertheless, since the key parameters of both the commercial silicon SPDs are balanced aiming for the general-purpose use, in some specific applications the parameter of most
interest should be individually optimized with the cost of the degradation of other parameters. For instance, in multiphoton entanglement experiments~\cite{RMP}, PDE is the most important parameter. Improving PDE of SPDs can directly increase the intensity of multiphoton entanglement and thus shorten the experimental integration time, while the accompanying increase of DCR almost does not affect its visibility due to the characteristic of multiple-fold coincidence.

\begin{figure*}[tbp]
\centering
\includegraphics[width=16 cm]{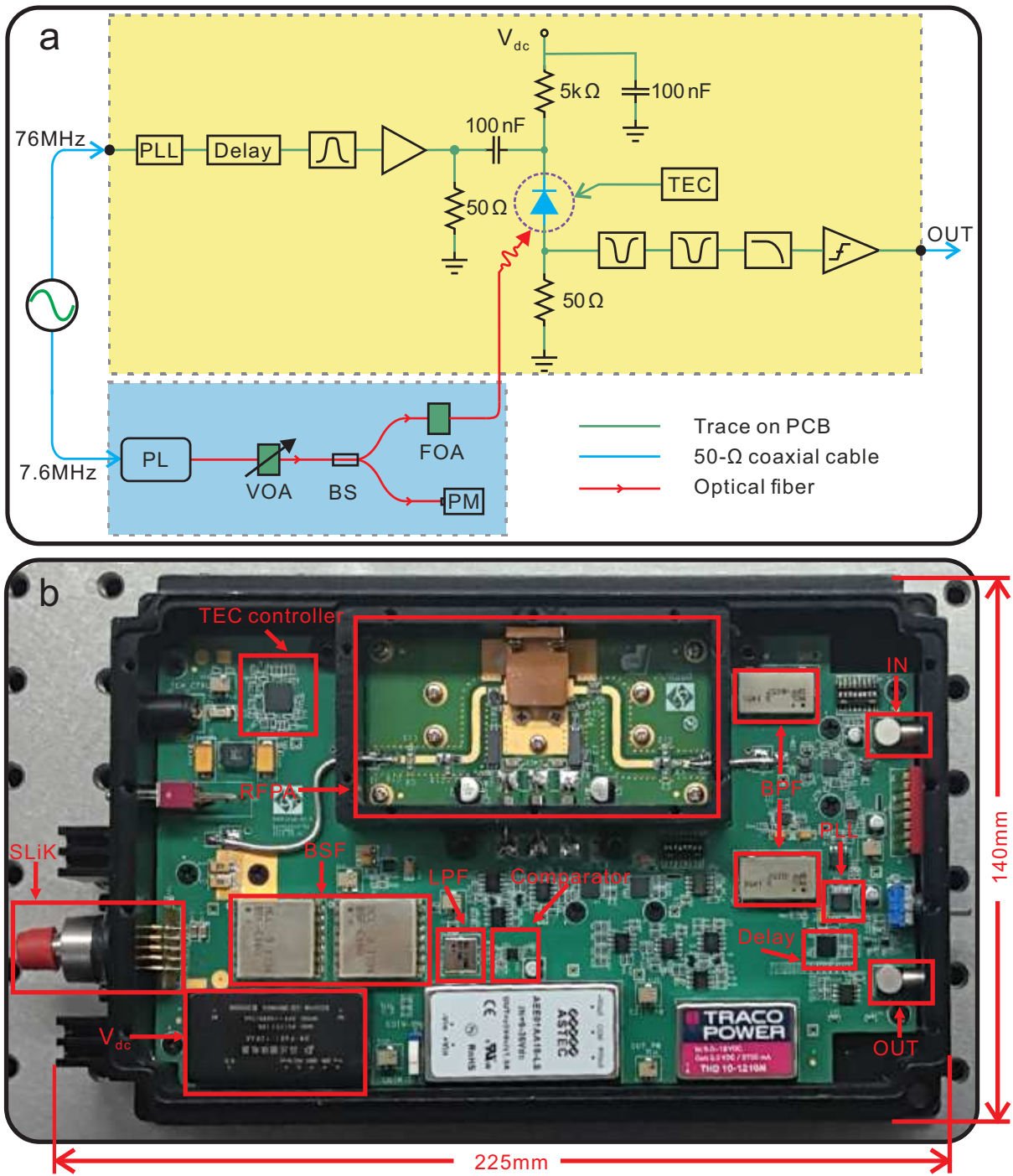}
\caption{Schematic diagram and the experimental setup for single-photon characterization at 785 nm (a) as well as the photo (b) of sine wave gating silicon SPD.
PLL, phase-locked loop;
Delay, tunable delay module;
BPF, band-pass filter;
RFPA, radio frequency power amplifier;
$V_{dc}$, DC bias voltage;
TEC, thermoelectric cooler;
BSF, band-stop filter;
LPF, low-pass filter;
PL, pulsed laser;
VOA, variable optical attenuator;
BS, beam splitter;
FOA, fixed optical attenuator;
PM, power meter.
\label{fig1}}
\end{figure*}

Further, given $n$ down-converted photon pairs created from a pulsed laser to generate multiphoton entangled states, the $2n$-fold coincidence count rate $C$ can be simply estimated as
\begin{equation}
\label{c}
C \sim f\mu ^{n}\eta ^{2n},
\end{equation}
where $f$ is the repetition frequency of pump laser, $\mu$ is the probability of generating one photon pair per pulse, and $\eta$ is the overall efficiency of each channel including the optical coupling efficiency and the detection efficiency of SPD (for simplicity, $\eta$ is assumed to be the same in each channel).
When PDE is improved by only 5\% (relative value), the coincidence count rate can be increased by 63\% for ten-photon entanglement experiment~\cite{Ten1,Ten2}, which implies that the experimental acquisition time can be greatly saved.

The PDE of silicon SPD primarily depends on excess bias voltage $V_{ex}$ ($V_{ex}=V_b-V_{br}$, where $V_b$ is the bias voltage and $V_{br}$ is the breakdown voltage of SPAD.
For 
SLiKs~\cite{SPCM} with considerably high $V_{br}$, $V_{ex}$ is normally larger than 30 V to achieve high avalanche probability. Given the same 
SLiK, in order to obtain higher PDE than SPCMs, $V_{ex}$ should be further increased, which brings a technical challenge to redesign the quenching electronics.

In this paper, we present sine wave gating (SWG) silicon SPDs dedicated to the application of multiphoton entanglement.
The SWG scheme is well suited for synchronous single-photon detection~\cite{Zhang15}.
The quenching electronics based on high-frequency sine wave gating~\cite{Zhang15,NSI06,GAP09,NAI09,GAP10,NTY11,LLW12,GAP12,RBM13,SNT14} are applied to the 
SLiKs disassembled from the commercial products of SPCMs. After the replacement, the PDE of each SPD is improved by 4 $\sim$ 9\% compared with its original performance at a wavelength of 785 nm. We then perform a four-photon entanglement experiment using such SWG SPDs. The count rate of four-photon entangled states is increased by 30\% without degrading its visibility compared with the result using SPCMs.

\section{Silicon SPD design}

\begin{figure}[tbp]
\centering
\includegraphics[width=8 cm]{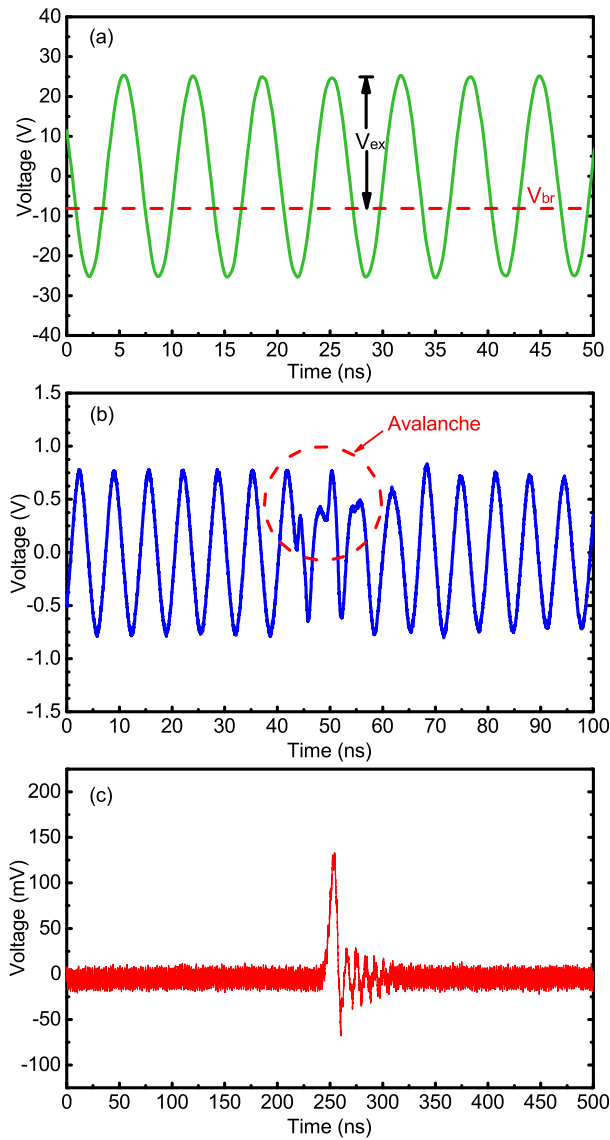}
\caption{Signal waveforms captured in an oscilloscope including (a) 152 MHz sine gates of 50 V, (b) avalanche signal superimposed with capacitive response signal with $V_{ex}$ of 32 V, and (c) extracted avalanche signal after filtering circuits.
\label{fig2}}
\end{figure}

Fig.~\ref{fig1} shows the schematic diagram and the photo of the sine wave gating silicon SPD.
The synchronized signal from the mode-locked Ti: sapphire laser with a repetition frequency of 76 MHz is used as the input of SPD.
The input signal is doubled to 152 MHz first by a phase-locked loop (PLL) in order to shorten the gating width and thus to suppress the afterpulsing effect.
Due to the limits of SPAD responses and weak avalanche extraction, the gate frequency cannot be too high.
A tunable delay module is used to adjust the delay between the gates and photon arrivals in order to maximize PDE.
The digital gate signals are converted into sine waves by a band-pass filter (BPF), and then the sine gates are amplified up to 50 V of $V_{\text{pp}}$
with a narrow-band radio frequency power amplifier (RFPA). The large-amplitude sine gates are AC coupled to the cathode of 
SLiK that is disassembled from a
commercial SPCM. Such amplitude can apply higher $V_{ex}$ to the 
SLiK than SPCM, to achieve potentially higher PDE, as shown in Fig.~\ref{fig2}(a).
The whole SPD system is supplied by a single 12 V DC power source with a total power dissipation of 18 W roughly.
Temperature is an important operation condition for 
SLiK.
Since the metal box of SWG SPD is also used as a heat sink for both SLiK and the whole quenching electronics, particularly RFPA,
the stable operation temperature of SLiK cannot reach below $\sim$ 260 K. In our SPD,
this temperature is finally set as 270 K for continuous operation using an integrated thermoelectric cooler (TEC).

The output signal from the 
SLiK is sampled on a 50-$\Omega$ resistor, and the avalanche signal is superimposed with the capacitive response signal, as shown in
Fig.~\ref{fig2}(b). Further, this mixed signal passes through two band-stop filters (BSF, BSF-C160+, Mini-Circuits) with a stop band frequency from 150 MHz to 170 MHz and a rejection of 48 dB and one low-pass filter (LPF, RLP-105+, Mini-Circuits) with a cut-off frequency of 116 MHz to eliminate the capacitive response signal and to extract the avalanche signal.
After the filtering circuits, the final avalanche signal, as shown in Fig.~\ref{fig2}(c), can be easily discriminated by a comparator.

\section{SPD characterization and application to four-photon entanglement experiment}

We perform single-photon characterization at 785 nm for 6 
SLiKs
disassembled from commercial SPCMs, and performance comparison between our SWG SPDs and the original SPCMs.
The single-photon characterization setup is shown in Fig.~\ref{fig1}(a).
The pulsed laser (PL, picoquant) is triggered by a 7.6 MHz synchronized LVTTL signal,
and the laser pulses with a width below 100 ps pass through a variable optical attenuator (VOA), a 99:1 beam splitter (BS), and a fixed optical attenuator (FOA).
The laser intensity is finally attenuated down to a level with a mean photon number per pulse of 0.1.
One port of the BS is continuously monitored by a calibrated optical power meter (PM).
We then use a counter (53220A, Agilent) and a time-to-digital converter (U1051A, Agilent) with 50 ps timing resolution
to measure the parameters of PDE, DCR and afterpulse probability ($P_{ap}$) for our SPDs and the original SPCMs.

\begin{table*}[tbp]
\footnotesize
\caption{
Performance comparison between SPCMs and sine wave gating (SWG) SPDs for the same 
SLiKs. SPCMs are characterized first, and then the disassembled 
SLiKs are mounted in SWG SPDs for corresponding characterization. $\Delta\eta$ represents the relative value of PDE improvement, defined as (PDE$^{SWG}$-PDE$^{SPCM}$)/PDE$^{SPCM}$.
\label{tab1}}
\begin{threeparttable}
\renewcommand{\arraystretch}{1.4}
\begin{tabularx}{15 cm}{@{\extracolsep{\fill}}lccccccc}
\hline
\hline
\multirow{2}{*}{Part number}    &\multirow{2}{*}{PDE$^{SPCM}$}   &\multirow{2}{*}{PDE$^{SWG}$}   &DCR$^{SPCM}$   &DCR$^{SWG}$  &\multirow{2}{*}{$P_{ap}$$^{SPCM}$}   &\multirow{2}{*}{$P_{ap}$$^{SWG}$}   &\multirow{2}{*}{$\Delta\eta$}\\& & &(cps) &(cps)\\
\hline
SPCM CD 3428 H 24276       &0.694          &0.757		   &360          &410          &0.005              &0.006              &0.09\\
SPCM CD 3428 H 24421       &0.718          &0.776          &230	    	 &206          &0.026              &0.007              &0.08\\
SPCM CD 3351 H 19730       &0.666          &0.697          &2091         &914          &0.023              &0.037              &0.05\\
SPCM-AQRH-13-FC 20800      &0.678          &0.708          &160          &68           &0.013              &0.011              &0.04\\
SPCM CD 3351 H 19695       &0.681          &0.729          &210          &126          &0.022              &0.038              &0.07\\
SPCM-AQRH-13-FC 15765      &0.680          &0.721          &3043         &1350         &0.017              &0.015              &0.06\\
\hline
\hline
\end{tabularx}
\end{threeparttable}
\end{table*}

The measured results are shown in Table~\ref{tab1}. It clearly shows that the parameter of most
interest, PDE, has been relatively increased in a range from 4\% to 9\% for each commercial SPCM.
Moreover, DCR performance of SWG SPDs is basically lower than that of SPCMs, due to the advantage of duty cycle in our quenching scheme.
For one 
SLiK (part number: SPCM CD 3428 H 24276), DCR of SWG SPD is slightly higher than SPCM, since in this case $V_{ex}$ is set to an extremely high value to obtain higher PDE.
As for the parameter of $P_{ap}$, the results in two cases are comparable each other. The SWG SPD cannot show significant advantage over SPCM in terms of afterpulse probability,
which is due to the fact that $P_{ap}$ is related with many factors such as operation temperature, excess bias voltage, hold-off time, avalanche duration time and parasitic capacitance.

\begin{figure}[tbp]
\centering
\includegraphics[width=8 cm]{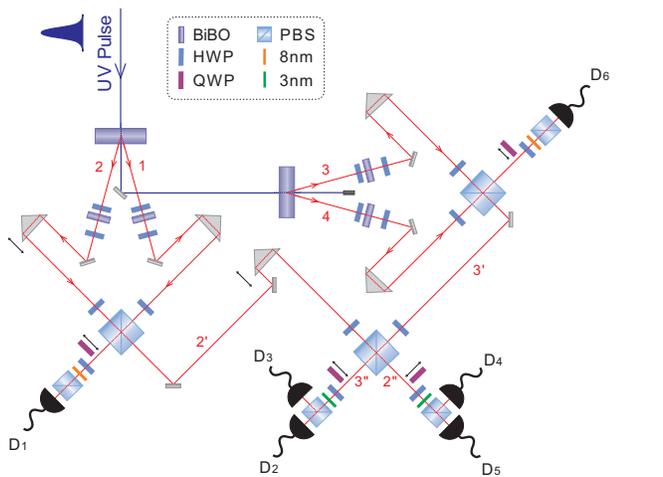}
\caption{Experimental setup for the generation of four-photon GHZ state.
An ultrafast pump laser with a central wavelength of 390~nm and a FWHM of 2.1~nm successively passes through
two BiBO crystals to generate polarization-entangled photon pairs.
The polarization of each output photon is analyzed using a combination of
a QWP, a HWP, and a PBS, together with a 
SPD in each output of the PBS.
Bandpass filters with a FWHM of 3 nm are used
on path 2 and 3 to erase the time information between the two entangled photon pairs~\cite{Grice01}.
Bandpass filters with a FWHM of 8 nm are used on path 1 and 4.
QWP: quarter-wave plate; HWP: half-wave plate; PBS: polarizing beam splitter.
\label{fig3}}
\end{figure}

These SWG SPDs are then used for the observation of multiphoton entangled states
in order to evaluate the increase of coincidence count rate. The experimental setup is shown in Fig.~\ref{fig3}.
An ultraviolet laser successively passes through two BiB$_{3}$O$_{6}$ (BiBO) crystals
to independently create two polarization entangled photon pairs via spontaneous parametric down conversion (SPDC) process, as illustrated in Fig.~\ref{fig3}.
Since we apply the technique of Bell-state synthesizer~\cite{Kim03} to prepare the entangled photon pairs, these two polarization entangled states can be written as
\begin{eqnarray}
\label{epr}
  |\Phi \rangle_{1} &=& \frac{1}{\sqrt{2}}(|H\rangle_{1}|H\rangle_{2}+|V\rangle_{1}|V\rangle_{2}), \\
  |\Phi \rangle_{2} &=& \frac{1}{\sqrt{2}}(|H\rangle_{3}|H\rangle_{4}+|V\rangle_{3}|V\rangle_{4}),
\end{eqnarray}
where $|H\rangle$ ($|V\rangle$) represents horizontal (vertical) polarization.

Photon 2 and photon 3 are then directed to a polarizing beam splitter (PBS) to
ensure spatial indistinguishability between photons from the two different BiBO crystals.
Through fine adjustments, photon 2 and 3 simultaneously arrive at the PBS.
Since the PBS transmits horizontally polarized photons and reflects vertically polarized photons,
the four-fold coincidence events of photon 1, 2, 3 and 4 indicate
the demonstration of four-photon Greenberger-Horne-Zeilinger (GHZ) state
\begin{eqnarray}
\label{4photon}
|\Psi \rangle_{12'3'4}= \frac{1}{\sqrt{2}}(|H\rangle_{1}|H\rangle_{2'}|H\rangle_{3'}|H\rangle_{4} \nonumber \\
+|V\rangle_{1}|V\rangle_{2'}|V\rangle_{3'}|V\rangle_{4}).
\end{eqnarray}

During the measurements, we use the Hong-Ou-Mandel-type interference~\cite{HOM87} to
evaluate the performance of prepared four-photon GHZ states.
Each photon is measured in the basis of $|\pm\rangle$
($|\pm\rangle = (|H\rangle\pm|V\rangle)/\sqrt{2}$),
which can be realized by setting the HWP at $22.5^{\circ}$.
In the experiment, we measure photons 1 and photon 4 deterministically in the $|+\rangle$ basis.
Theoretically, the four-photon GHZ state only contributes to the components of $|++++\rangle$ and $|+--+\rangle$,
that is, only two four-fold coincidence counts, D1D2D5D6 and D1D3D4D6, can register.
However, in practical situation, the higher-order emissions and
partial distinguishability between SPDC photons can yield the noisy terms, $|++-+\rangle$ and $|+-++\rangle$.
Thus, the four-fold coincidence count events contain all four components,
i.e., $|++++\rangle$, $|+-++\rangle$, $|++-+\rangle$, and $|+--+\rangle$.

\begin{figure}[tbp]
  \centering
  \includegraphics[width=8 cm]{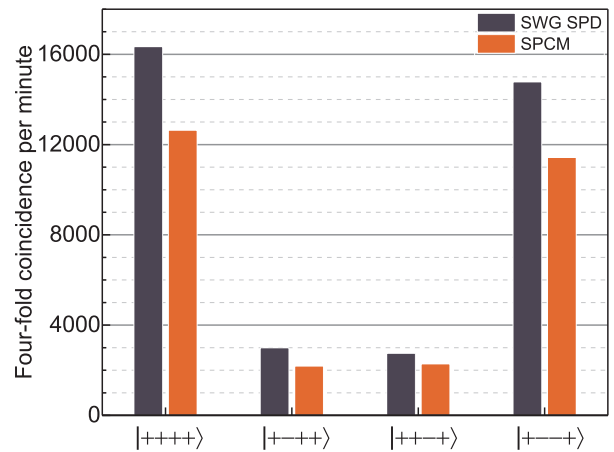}
  \caption{The measured results of four-fold coincidence counts per minute in the basis of $|\pm\rangle$ using SWG SPDs
and corresponding SPCMs, respectively.
\label{fig4}}
\end{figure}

The measured results of four-fold coincidence counts per minute using both SWG SPDs
and the corresponding SPCMs are listed in Fig.~\ref{fig4}, in which the coincidence count rate of four-photon
entangled states is increased by 30\%. Such increase agrees very well with the values of $\Delta\eta$ as shown in Table~\ref{tab1},
and most importantly the visibility of entangled photons is not degraded, with 68.7\% for SPCMs and 68.9\% for SWG SPDs.

\section{conclusion}

In conclusion, we have presented high detection efficiency silicon SPDs aiming for multiphoton entanglement experiments
based on the SWG scheme. After redesigning the quenching electronics for 
SLiKs disassembled from commercial SPCMs,
the characterized performance, particularly PDE, can be effectively improved. These SWG SPDs have been used in four-photon entanglement
experiment, and the four-fold coincidence count rate is increased by 30\% without degrading its visibility compared with the original SPCMs.
Our approach provides an effective solution for the applications requiring ultrahigh detection efficiency silicon SPDs.

\section{acknowledgement}

This work has been financially supported by the National Basic Research Program of China Grant No.~2013CB336800, the National Natural Science Foundation of China Grant No.~11674307, and the Chinese Academy of Sciences. N. Zhou, and W.-H. Jiang contributed equally to this work.

\end{document}